\documentclass[aps,prl,twocolumn,floatfix]{revtex4-2}

\usepackage{bm}
\usepackage{mathrsfs,amsbsy,amssymb,latexsym,amsfonts,amsmath}
\usepackage{graphicx}
\usepackage{color}
\usepackage{url}
\usepackage[hidelinks]{hyperref}

\newcommand{\bv}[1]{{\boldsymbol #1}}

\newcommand{\kB}{k_\mathrm{B}}
\newcommand{\nm}{\nonumber\\}
\newcommand{\eff}{\mathrm{eff}}

\newcommand{\ep}{\varepsilon}
\newcommand{\Ps}{p_\mathrm{s}}

\newcommand{\Tm}{T_\mathrm{m}}

\begin{document}

\title{Effective Gravity Governs Liquid--Gas Inversion under Heat Conduction}

\author{Satoshi Uruno}
\affiliation{Department of Physics, Ibaraki University, Mito 310-8512, Japan}
\author{Naoko Nakagawa}
\email{naoko.nakagawa.phys@vc.ibaraki.ac.jp}
\affiliation{Department of Physics, Ibaraki University, Mito 310-8512, Japan}

\date{\today}

\begin{abstract}
Using standard molecular dynamics of a two-dimensional Lennard--Jones fluid, we demonstrate that heat conduction changes the liquid--gas arrangement as if gravity were shifted.
Under weak heat conduction with the lower boundary hotter than the upper boundary, the transition between liquid below gas and liquid above gas shifts from zero to finite downward gravity.
Global thermodynamics predicts that the equilibrium saturation-pressure difference between the two boundary temperatures acts as a correction opposing gravity, defining an effective gravity that selects the liquid--gas arrangement.
The observed transition occurs at zero effective gravity, where this difference balances the hydrostatic pressure difference, in agreement with the prediction based on independently measured saturation pressures.
The hysteresis response under heat conduction also agrees with the equilibrium response when plotted against effective gravity.
These results establish global thermodynamics as a predictive method for the macroscopic organization of a molecular fluid maintained away from equilibrium by heat conduction.
\end{abstract}

\maketitle

Gravity places a liquid below its gas, but heat conduction can invert this arrangement.
Can thermodynamics predict when the inversion occurs?
Nonuniform fluids are conventionally described by local hydrodynamic fields, density profiles, and interfacial free energies \cite{LandauLifshitzFluid,RowlinsonWidom,BLS65,Weeks84}.
Under heat conduction, continuum hydrodynamics \cite{Onuki07} and molecular simulations \cite{Ogushi06,Muscatello17} describe interfacial structure and heat transport in liquid--gas coexistence.
Gravity selects the vertical arrangement, suppresses interface fluctuations, and shifts local coexistence conditions through hydrostatic pressure variation \cite{BerberanSantos02,Sikkenk87,Czelusniak23}.
Thermal driving can nevertheless place a denser phase above a lighter one in weakly first-order transitions \cite{Busse71,Ahlers93,Sakurai99}.
Molecular dynamics has also revealed a liquid layer hovering against gravity under heat conduction without convection \cite{YNS24}.
Deciding whether liquid is thermodynamically favored below or above gas, however, requires comparing the two configurations as states of the whole system; in equilibrium, this comparison is supplied by their free energies \cite{Callen}.
For fluids maintained away from equilibrium by a heat current, no comparably general thermodynamic selection principle is established.

Global thermodynamics extends the whole-system comparison to weak heat conduction and gravity \cite{NS17,NS19,NS22,NSgravity25}.
Its predictions for nonequilibrium phase coexistence have been supported by Hamiltonian Potts simulations \cite{KNS23} and by theoretical and numerical studies of boundary-driven diffusion \cite{SN25}.
For planar liquid--gas configurations, the theory compares bulk thermodynamic states without an explicit surface-free-energy term.
In the isothermal limit, global thermodynamics identifies the change between the two arrangements at $g=0$ as a first-order thermodynamic configurational transition, distinct from the ordinary liquid--gas transition \cite{NSgravity25}.
The corresponding isothermal center-of-mass response, supported by equilibrium molecular dynamics \cite{NSgravity25}, provides the equilibrium reference for comparison with heat conduction.
Under heat conduction, global thermodynamics predicts that the sign of an effective gravity $g_\eff$ selects whether liquid lies below or above gas \cite{NSgeff26,NSfull26}.
Heat conduction leaves the gravitational force unchanged, but the difference in equilibrium saturation pressure between the two boundary temperatures acts as a correction to gravity in selecting the liquid--gas arrangement.
The thermodynamically favored arrangement is predicted to change at $g_\eff=0$, rather than at $g=0$.

In this Letter, we demonstrate that the effective-gravity selection principle predicts the liquid--gas arrangements realized by standard molecular dynamics.
We show that weak heat conduction shifts the configurational transition from $g=0$ to positive gravity by the predicted amount, so that liquid remains at the cold upper wall against gravity.
The hysteresis response under heat conduction also agrees with the equilibrium response when expressed in terms of effective gravity.
The predicted arrangements emerge from Newtonian bulk dynamics with gravity and boundary heat baths, establishing the thermodynamic selection principle as a predictive method for a molecular fluid.

\paragraph*{Effective gravity.---}
Let the lower and upper baths have temperatures $T_1$ and $T_2$, respectively, with $\Delta T\equiv T_1-T_2\geq0$; $g>0$ denotes downward gravity.
Let $\Ps(T)$ denote the equilibrium saturation pressure, and define the saturation-pressure difference between the boundary temperatures as $\Delta\Ps\equiv\Ps(T_1)-\Ps(T_2)$.
We quantify the departure from equilibrium by the relative saturation-pressure difference,
$\ep_p\equiv\Delta\Ps/\Ps(\Tm)$, where $\Tm\equiv(T_1+T_2)/2$.
The effective gravity predicted by global thermodynamics \cite{NSgeff26,NSfull26} is
\begin{align}
 g_\eff\equiv g-\frac{\Delta\Ps}{m\bar\rho L},
 \label{eq:geff}
\end{align}
where $m$ is the particle mass, $\bar\rho$ is the mean number density, and $L$ is the cell height.
The hydrostatic pressure difference between the lower and upper boundaries (lower minus upper) is $\Delta p\equiv m\bar\rho Lg$.
This pressure difference follows from the local force balance integrated over the cell, which is why the mean density $\bar\rho$ appears rather than the density of either phase alone.

The condition $g_\eff=0$ is the predicted thermodynamic transition point between the two configurations \cite{NSgeff26}: $g_\eff>0$ selects liquid below gas, whereas $g_\eff<0$ selects liquid above gas.
The prediction \eqref{eq:geff} applies for weak gravity and weak heat conduction, where corrections of order $\ep_p^2$ are negligible.
In terms of $\Delta p$, the transition condition can be written as a balance between the hydrostatic and saturation-pressure differences,
\begin{align}
 g_\eff=0\quad\Longleftrightarrow\quad\Delta p=\Delta\Ps.
 \label{eq:pressure-balance}
\end{align}
For $\Delta T>0$, the transition is thus shifted from its equilibrium position $g=0$ to $g=\Delta\Ps/(m\bar\rho L)>0$, allowing liquid to be favored above gas against gravity.

\paragraph*{Simulation setup.---}
To test the prediction \eqref{eq:geff}, we perform standard classical molecular dynamics for $N$ particles of a two-dimensional Lennard--Jones fluid in a rectangular cell of height $L$ and horizontal width $W$ under gravity.
We use Lennard--Jones reduced units, $m=\sigma=\epsilon=\kB=1$, where $\sigma$ and $\epsilon$ are the length and energy scales of the particle interaction, respectively, and time is measured in $\sigma\sqrt{m/\epsilon}$.
The bulk evolves under Newtonian dynamics.
Periodic boundary conditions are imposed in the horizontal direction, while the lower and upper boundaries are attractive Lennard--Jones walls; only particles within $8\sigma$ of the walls are coupled to Langevin baths at $T_1$ and $T_2$, respectively [Fig.~\ref{fig:opening}(a)].
For the interaction potentials and full cell dimensions, see Appendix~\ref{sec:model}.

Figures~\ref{fig:opening}(b) and \ref{fig:opening}(c) show representative configurations at the same positive $g$: liquid lies below gas in equilibrium but above it under heat conduction.
To quantify whether the liquid lies below or above the gas, we measure the vertical center of mass of all particles,
\begin{align}
 X\equiv\frac{1}{N}\sum_{i=1}^{N}x_i,
 \label{eq:X}
\end{align}
where $x_i$ is the vertical coordinate of particle $i$, measured upward from the lower wall at $x=0$.
We denote the time-averaged center of mass by $\langle X\rangle$.
We use the scaled center of mass $X/L$ as the configurational order parameter.
Values below and above $1/2$ indicate liquid below and above gas, respectively.

\begin{figure}[t]
\centering
\includegraphics[width=\linewidth]{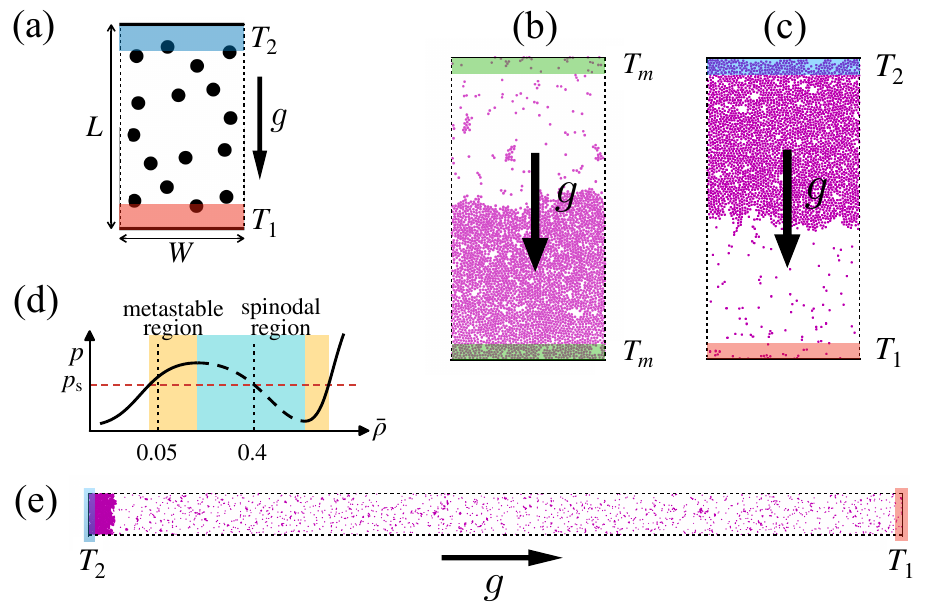}
\caption{System and representative separated configurations at $\Tm=0.43$.
(a) Simulation geometry.
(b),(c) Equilibrium and heat-conducting configurations, respectively, for $(\bar\rho,N,L/W)=(0.4,3200,2)$ at $g=1.32\times10^{-5}$; $\ep_p=0$ in (b) and $\ep_p\simeq0.09$ in (c), with $T_1=0.4325$ and $T_2=0.4275$.
(d) Schematic isotherm. The vertical dotted lines mark the two mean densities used in the simulations, $\bar\rho=0.05$ and $0.4$; the horizontal dashed line denotes $\Ps$.
Yellow and cyan indicate the metastable and spinodal regions, respectively.
(e) The elongated near-binodal system, $(\bar\rho,N,L/W)=(0.05,3380,20)$, at $\Delta T=0.02$ ($\ep_p\simeq0.36$), displayed horizontally with gravity pointing right and heat flowing left.}
\label{fig:opening}
\end{figure}

The saturation pressure $\Ps(T)$ is calibrated independently in equilibrium simulations, as described in Appendix~\ref{sec:pressure}.
We fix $\Tm=0.43$ and vary $\Delta T$ and $g$.
At $\Tm=0.43$, the saturation densities of gas and liquid are estimated as $\rho_{\rm c}^{\rm G}\simeq0.036$ and $\rho_{\rm c}^{\rm L}\simeq0.73$.
We choose two systems with $(\bar\rho,N,L/W)=(0.05,3380,20)$ and $(0.4,3200,2)$, keeping $N$ and $W$ nearly fixed while changing mainly $L$.
Their liquid volume fractions are about $0.02$ and $0.53$, respectively.
As shown in Fig.~\ref{fig:opening}(d), the two systems lie near the gas-side binodal and in the coexistence interior, respectively.
Even the near-binodal system forms a planar liquid layer about $20\sigma$ thick, as shown in Fig.~\ref{fig:opening}(e).

Because $N/W=\bar\rho L$ is nearly unchanged between the two systems, the hydrostatic pressure difference $m\bar\rho Lg$ is nearly the same at a common $g$.
The predicted gravity shift $\Delta\Ps/(m\bar\rho L)$ in \eqref{eq:geff} is therefore also nearly the same for the two systems at a common temperature difference.
Global thermodynamics leads us to expect little or no configurational hysteresis for the first system but clear hysteresis for the second \cite{NSgravity25,NSfull26}.
With these two systems, we test whether the equilibrium and heat-conduction values of $\langle X\rangle/L$ agree when compared at the same $g_\eff$, both with and without pronounced hysteresis.

Movies S1 and S2 in the Supplemental Material \cite{SM} show the response of the two initially isothermal systems to an imposed temperature difference.
Under unchanged downward gravity and without convective flow, liquid disappears at the heated lower wall and forms at the cooled upper wall.
This reorganization differs qualitatively from gravitational settling.
Yet global thermodynamics predicts liquid above gas in both movies: the imposed temperature difference makes the saturation-pressure difference exceed the hydrostatic pressure difference, giving $g_\eff<0$ in \eqref{eq:geff}.

\paragraph*{Heat-induced shift of the configurational transition.---}
We first examine the system at $\bar\rho=0.05$, near the gas-side binodal in Fig.~\ref{fig:opening}(d).
To test the predicted transition at $g_\eff=0$, we compare equilibrium with heat conduction at $\Delta T=0.02$.
We locate the transition by scanning $g$ in both directions, which also tests for hysteresis.
Increasing scans start with liquid above gas at negative $g$, and decreasing scans start with liquid below gas at positive $g$.
We change $g$ stepwise without reinitializing the configuration between neighboring values.
The principal spacing is $\delta g=2\times10^{-5}$.
The holding time at each $g$ is $2.5\times10^6$ Lennard--Jones time units, about five times the observed density-profile relaxation time; this relaxation time is distinct from the waiting time for a configurational switch.
We examine the transition using $\langle X\rangle/L$ at each $g$, with the average taken over the latter half of the holding time.
For further details of the scanning and averaging procedures, see Appendix~\ref{sec:protocol}.

Figure~\ref{fig:dt0}(a) compares $\langle X\rangle/L$ at $\Delta T=0$ and $0.02$ as functions of $g$.
In equilibrium, liquid lies above the gas for $g<0$ and below it for $g>0$.
The increasing and decreasing scans change configuration near $g=0$, with no resolved hysteresis width.
Under heat conduction, the transition shifts to positive $g$ in both directions.
Liquid therefore remains above gas over a finite range of downward gravity for which it lies below gas in equilibrium.

Using the independently measured equilibrium saturation pressures, we predict a shift $\Delta\Ps/(m\bar\rho L)\simeq7\times10^{-5}$ at $\Delta T=0.02$, appreciably larger than the spacing between sampled $g$ values.
To compare the two responses at equal effective gravity, we plot the same data against $g_\eff$ in Fig.~\ref{fig:dt0}(b), with $g_\eff=g$ for equilibrium.
Both transitions now occur near $g_\eff=0$ within the resolution of the $g$ scan.
The values of $\langle X\rangle/L$ agree with those in equilibrium on both sides of the transition as well.
The equilibrium saturation-pressure difference $\Delta\Ps$ thus predicts the observed shift of the configurational transition.

\begin{figure}[t]
\centering
\includegraphics[width=0.95\linewidth]{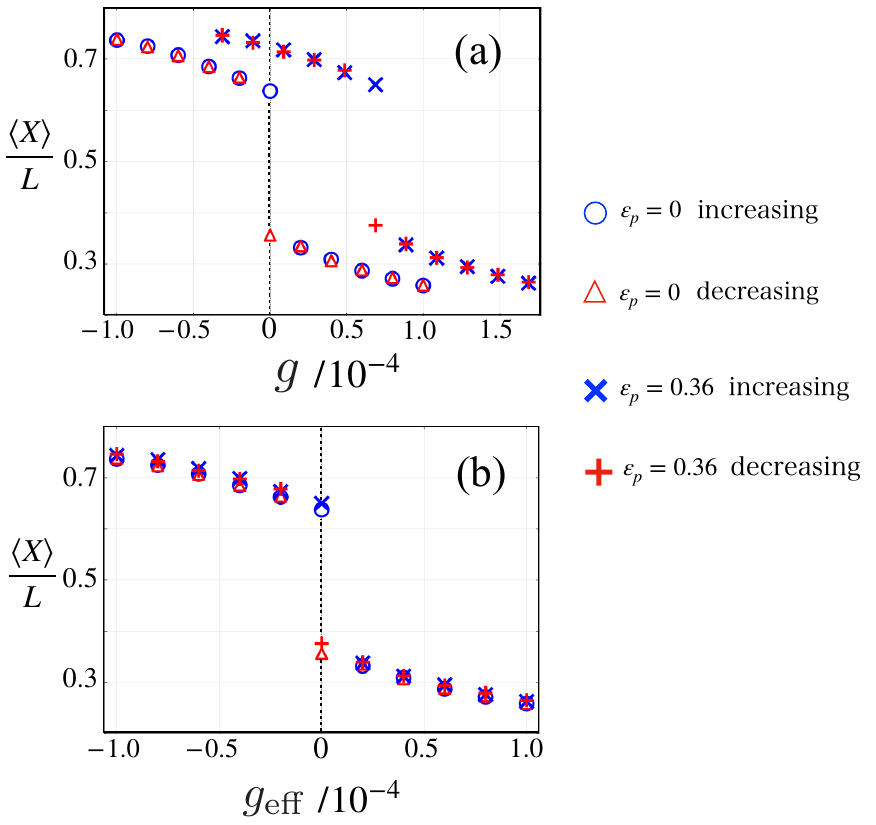}
\caption{Effective-gravity test near the gas-side binodal, $(\bar\rho,N,L/W)=(0.05,3380,20)$.
The same representative $\langle X\rangle/L$ responses are plotted against (a) $g$ and (b) $g_\eff$.
Open circles and triangles denote increasing and decreasing scans at $\Delta T=0$; crosses and pluses denote the corresponding scans at $\Delta T=0.02$ ($\ep_p\simeq0.36$).}
\label{fig:dt0}
\end{figure}

\begin{figure}[t]
\centering
\includegraphics[width=0.79\linewidth]{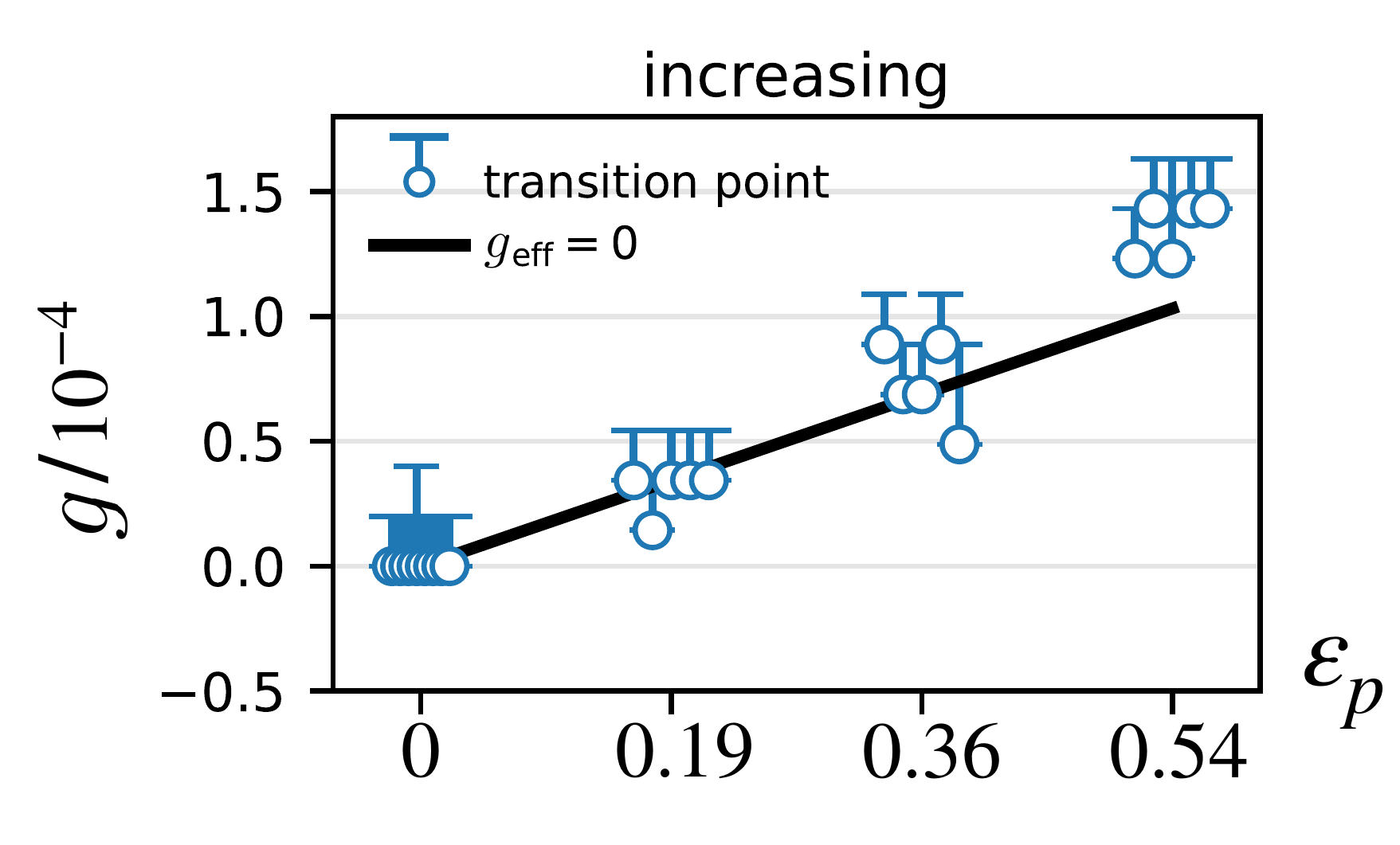}
\par\smallskip
\includegraphics[width=0.79\linewidth]{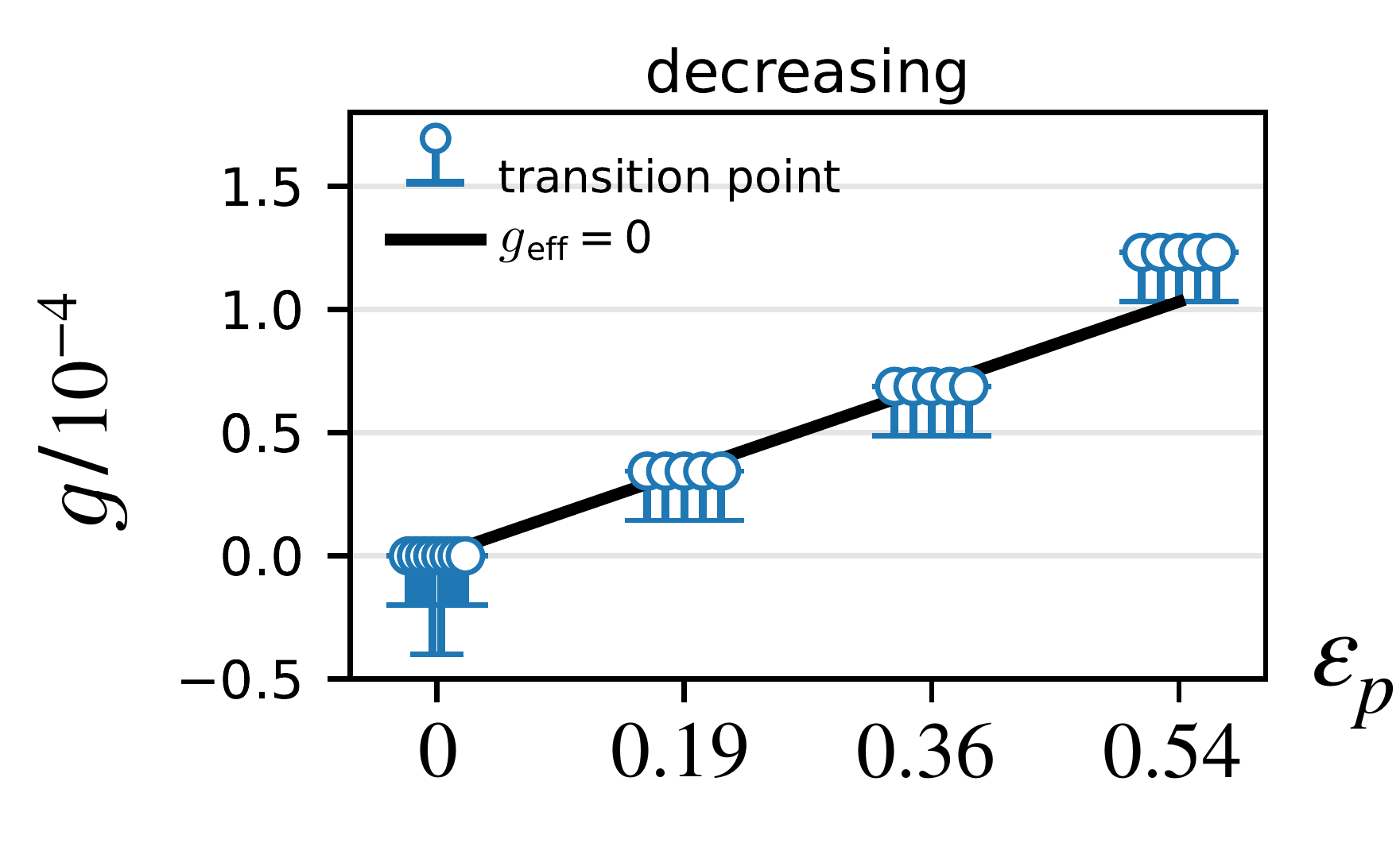}
\caption{Transition points for $L/W=20$ compared with the prediction $g_\eff=0$ (black line).
Upper and lower panels show increasing and decreasing $g$, respectively.
Circles mark the last pre-switch value; bars extend to the first post-switch value and show scan resolution, not statistical uncertainty.
Small horizontal offsets separate overlapping points at the same $\ep_p$.}
\label{fig:intervals}
\end{figure}

The predicted shift depends on $\Delta T$ through $\Delta\Ps$.
We test this dependence using independent runs at $\Delta T=0$, $0.01$, $0.02$, and $0.03$ in the same system.
We locate each transition from the change between configurations with late-time $X/L$ above and below $1/2$.
Figure~\ref{fig:intervals} plots these transition points against $\ep_p$.
The black line represents $g=\Delta\Ps/(m\bar\rho L)$, the predicted transition condition $g_\eff=0$.
At $\Delta T=0$, the points for both scan directions lie at $g=0$.
At $\Delta T=0.01$ and $0.02$ ($\ep_p\simeq0.19$ and $0.36$), the observed transition values shift to positive $g$ and remain close to the predicted line.
For increasing $g$ at $\Delta T=0.02$, some transition points lie about one $g$ increment above the line.
For these smaller temperature differences, the measured transition points agree with $g=\Delta\Ps/(m\bar\rho L)$ up to deviations on the scale of the $g$ spacing.

At $\Delta T=0.03$ ($\ep_p\simeq0.54$), the transition points lie above the predicted line in both directions, with a larger departure for increasing $g$.
Here $\ep_p$ is no longer small, and the departures are consistent with corrections beyond the leading-order prediction \eqref{eq:geff}.

\begin{figure}[t]
\centering
\includegraphics[width=\linewidth]{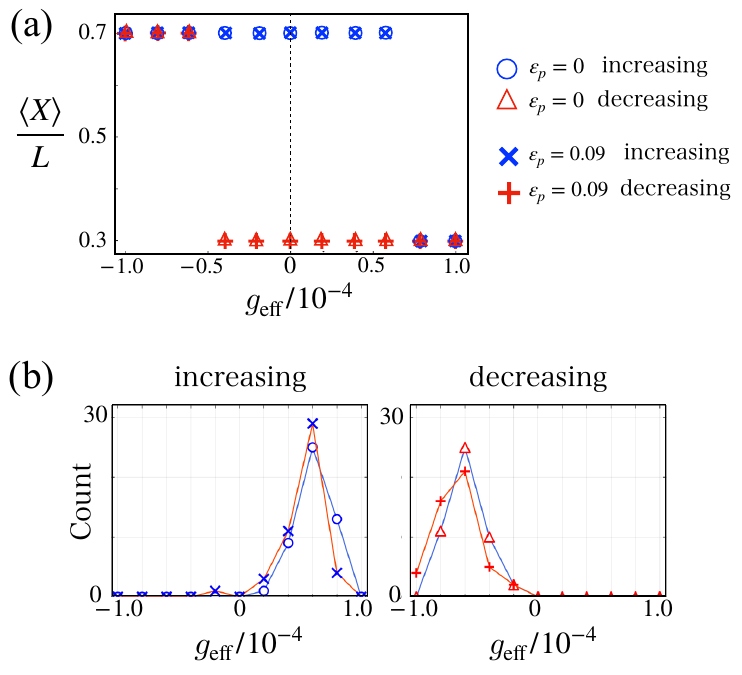}
\caption{Effective-gravity test in the hysteretic system, $(\bar\rho,N,L/W)=(0.4,3200,2)$.
(a) $\langle X\rangle/L$ responses from one selected cycle at each temperature difference.
(b) Switching-frequency distributions for increasing (left) and decreasing (right) $g$, each containing 48 events.
Blue (red) symbols denote increasing (decreasing) scans: open circles (triangles) for $\Delta T=0$ and crosses (pluses) for $\Delta T=0.005$ ($\ep_p\simeq0.09$).
Connecting lines in (b) are blue for $\Delta T=0$ and red for $\Delta T=0.005$ and guide the eye.}
\label{fig:ar2test}
\end{figure}

\paragraph*{Hysteretic regime in heat conduction.---}
We next turn to the system at $\bar\rho=0.4$ [Figs.~\ref{fig:opening}(b),(c)], in the coexistence interior shown in Fig.~\ref{fig:opening}(d).
We again compare $\langle X\rangle/L$ in equilibrium and under heat conduction on the $g_\eff$ axis.
At $\Delta T=0$ and $0.005$ ($\ep_p\simeq0.09$), we scan $g$ in both directions, holding each value for $5\times10^5$ Lennard--Jones time units, again about five density-profile relaxation times.
Figure~\ref{fig:ar2test}(a) shows the resulting responses, averaged as in Fig.~\ref{fig:dt0}.
In equilibrium, the liquid-above configuration persists into positive $g$ during an increasing scan, while the liquid-below configuration persists into negative $g$ during a decreasing scan.
The clearly separated switching values show pronounced configurational hysteresis: a configuration can remain observable after it ceases to be thermodynamically favored.
When plotted against $g_\eff$, the $\langle X\rangle/L$ values at $\Delta T=0.005$ agree with those at $\Delta T=0$ for both increasing and decreasing $g$.

To determine whether this agreement persists across repeated scans, we record the value of $g$ at which the configuration changes in each scan and express it in terms of $g_\eff$.
Figure~\ref{fig:ar2test}(b) compares the frequency distributions of these switching values at $\Delta T=0$ and $0.005$, separately for increasing and decreasing $g$.
The distribution peaks coincide at $g_\eff\simeq+6\times10^{-5}$ for increasing $g$ and $-6\times10^{-5}$ for decreasing $g$.
Heat conduction thus shifts the distribution peaks for both scan directions by the independently predicted amount.
Although individual switches need not occur at $g_\eff=0$, the agreement of the heat-conduction responses and switching distributions with the equilibrium hysteresis, whose thermodynamic transition is at $g=0$, supports the predicted transition condition.

Together, Figs.~\ref{fig:dt0}--\ref{fig:ar2test} show that replacing $g$ by $g_\eff$ accounts for both the transition shift and the hysteresis response under weak heat conduction.

\paragraph*{Discussion.---}
The two-dimensional molecular-dynamics simulations in Ref.~\cite{YNS24} used the same Lennard--Jones fluid and heat-bath setup as here, but with repulsive walls.
In the hovering state observed there, gas separates the liquid from both walls.
Replacing these repulsive walls with attractive walls allows liquid to contact either boundary, realizing the two-layer configurations compared by global thermodynamics.
Wall effects on the onset of phase formation have also been reported in lattice-Boltzmann simulations \cite{Czelusniak23}.
Under conditions comparable to those of Fig.~\ref{fig:opening}(c), the liquid between repulsive walls hovers in the lower part of the container, with $X/L<1/2$ \cite{YNS24}.
With attractive walls, liquid instead lies above gas at the cold upper wall [Fig.~\ref{fig:opening}(c)], and the observed transition shift agrees with the predicted condition $g_\eff=0$.

In ordinary liquid--gas transitions, metastable supercooled gas and superheated liquid sustain hysteresis \cite{Binder84}.
In configurational switching, global thermodynamics suggests an opposite role: near the binodal, a metastable gas or liquid filling the cell may facilitate exchange between the separated arrangements, whereas inside the spinodal the unstable single-phase state lies on a free-energy ridge between them \cite{NSgravity25,NSfull26}.
This offers a possible explanation for the contrasting hysteresis  in the two systems.
To leading order, the free-energy difference between this unstable state and either separated configuration takes its equilibrium form with $g$ replaced by $g_\eff$.
This provides a thermodynamic basis for comparing the hysteretic responses on the $g_\eff$ axis; for the free-energy differences, see Appendix~\ref{sec:landscape}.

At $\ep_p\simeq0.54$ ($\Delta T=0.03$), the nonequilibrium parameter is no longer small, and the increasing and decreasing scans depart asymmetrically from the effective-gravity prediction in Fig.~\ref{fig:intervals}.
Whether this reflects heat-induced changes in the transition barriers or switching dynamics beyond the thermodynamic description remains an open question \cite{NSfull26}.

For weak heat conduction, the observed transition shift and hysteresis response follow the equilibrium behavior with $g$ replaced by $g_\eff$.
The heat-induced loss of liquid at the hot wall and its formation at the cold wall differ from gravitational settling, yet the saturation-pressure difference enters the selection of the arrangement as a correction to gravity.
This microscopic realization establishes a thermodynamic route to predicting which macroscopic structures heat currents favor without resolving their formation processes.

\paragraph*{Acknowledgement.---}

The authors thank S. Sasa, A. Yoshida, R. Yamamoto, Y. Nakamura, F. Kagawa, and R. Higa for useful discussions.
The numerical simulations were performed with LAMMPS on the facilities of the Supercomputer Center, the Institute for Solid State Physics, the University of Tokyo.
This study was supported by JSPS KAKENHI Grant Numbers JP23K22415, JP25K00923, JP25K22002, and JP26H00383.

The numerical data were analyzed by the authors.
OpenAI Codex (GPT-6 Astra) and Anthropic Claude Code (Fable 5.1 and Opus 5.5) were used to assist with independent checks of these analyses, literature review, manuscript preparation, and the preparation of plotting code and formatting of figures under the authors' direction.
The authors reviewed and corrected the AI outputs and take full responsibility for the content and conclusions of this paper.

\paragraph*{Data Availability.---}

The data supporting the findings of this study are available from the corresponding author upon reasonable request.

\twocolumngrid
\appendix
\normalsize

\renewcommand{\theequation}{\Alph{section}\arabic{equation}}
\renewcommand{\theHequation}{\Alph{section}.\arabic{equation}}
\renewcommand{\thefigure}{\Alph{section}\arabic{figure}}
\renewcommand{\theHfigure}{\Alph{section}.\arabic{figure}}
\setcounter{section}{0}
\setcounter{equation}{0}
\setcounter{figure}{0}

\vspace{1cm}

\begin{center}
{\large\bfseries End Matter}
\end{center}

\refstepcounter{section}
\section*{Appendix \Alph{section}: Molecular-dynamics model}
\label{sec:model}

For the two-dimensional geometry in Fig.~\ref{fig:opening}(a), the horizontal boundary is periodic and the walls are at $x=0$ and $L$.
The Hamiltonian is
\begin{align}
 H={}&\sum_{i=1}^{N}\frac{|\bv p_i|^2}{2m}
 +\sum_{i<j}u(r_{ij})+\sum_{i=1}^{N}mgx_i \nm
 &+\sum_{i=1}^{N}\left[u(x_i)+u(L-x_i)\right].
 \label{eq:H}
\end{align}
Here $\bv p_i$ and $x_i$ are the momentum and vertical coordinate of particle $i$, and $r_{ij}$ is the distance between particles $i$ and $j$.
The particle--particle and wall--particle interactions use the same shifted and truncated 12--6 Lennard--Jones function,
\begin{align}
 u(r)=4\epsilon\left[\left(\dfrac{\sigma}{r}\right)^{12}
 -\left(\dfrac{\sigma}{r}\right)^6
 -\left(\dfrac{\sigma}{r_c}\right)^{12}
 +\left(\dfrac{\sigma}{r_c}\right)^6\right]
 \label{eq:lj}
\end{align}
for $r<r_c$, with a cutoff at $r_c=3\sigma$.

The lower and upper Langevin baths are held at
\begin{align}
 T_1=\Tm+\frac{\Delta T}{2},\qquad
 T_2=\Tm-\frac{\Delta T}{2}.
\end{align}
Particles within $8\sigma$ of each wall are coupled to the corresponding bath.
The damping time is $1.0$, and the integration time step is $0.005$, in the Lennard--Jones units defined in the main text.

The dimensions follow from $LW=N/\bar\rho$: $(W,L)\simeq(58.14,1163)$ for $(\bar\rho,N,L/W)=(0.05,3380,20)$ and $(63.25,126.5)$ for $(0.4,3200,2)$.

\refstepcounter{section}
\section*{Appendix \Alph{section}: Independent equilibrium pressure calibration}
\label{sec:pressure}
\setcounter{equation}{0}
\setcounter{figure}{0}

We determine $\Ps(T)$ from the whole-system two-dimensional virial pressure in equilibrium simulations at $g=0$, with both baths at temperature $T$.
We use $L/W=2$ and $\bar\rho=0.4$, well inside the coexistence region, and compare $N=5000$ with $N=3200$ to assess finite-size sensitivity.
After discarding the first $25\%$ of each record, we divide the remainder into 20 contiguous blocks.
We calculate the time-averaged pressure and estimate its standard error from the block means.

\begin{figure}[!t]
\centering
\includegraphics[width=1.5\columnwidth]{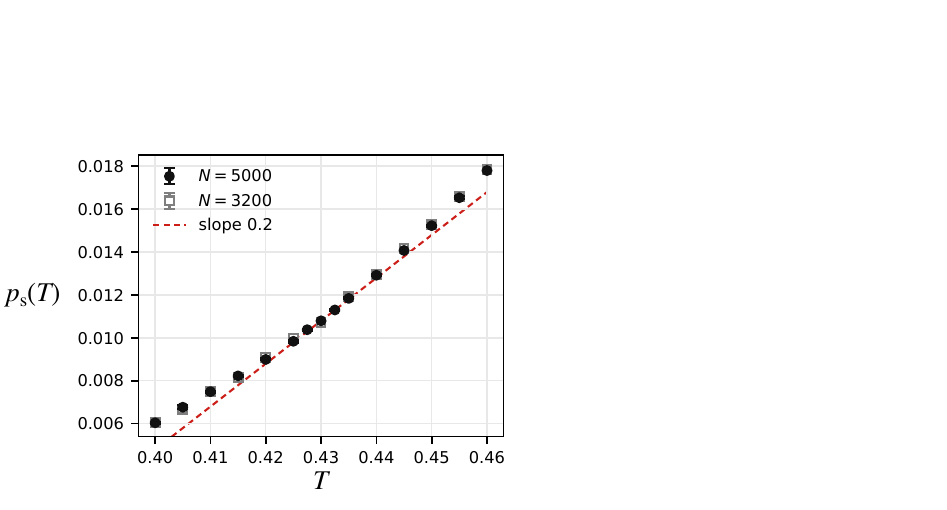}
\caption{Equilibrium virial-pressure calibration used to evaluate $\Delta\Ps$.
Filled circles and open squares show $N=5000$ and $3200$, respectively; bars are estimated standard errors.
The red dashed line is $\Ps(\Tm)+0.2(T-\Tm)$, through the measured $N=5000$ value at $\Tm=0.43$.}
\label{fig:pressure}
\end{figure}

In the thermodynamic limit, $\Ps(T)$ is a bulk coexistence property independent of $\bar\rho$ and $L/W$.
The two system sizes agree on the scale of Fig.~\ref{fig:pressure}.
We adopt the $N=5000$ series as a finite-system estimate of this bulk quantity for both geometries used in the main text.
For $\Delta T=0.005$, $0.01$, $0.02$, and $0.03$, the measured pressure differences $\Delta\Ps$ are $9.2\times10^{-4}$, $2.0\times10^{-3}$, $3.9\times10^{-3}$, and $5.8\times10^{-3}$, respectively.
We calculate $\ep_p$ from the measured $\Ps(T)$ throughout, including the nonlinear nonequilibrium regime at $\Delta T=0.03$.
The standard errors of these pressure differences, propagated to the predicted transition value $\Delta\Ps/(m\bar\rho L)$, are less than one tenth of the scan spacing $2\times10^{-5}$.

Near $\Tm=0.43$, the linear approximation in Fig.~\ref{fig:pressure} gives $\Delta\Ps\simeq0.2\Delta T$, which we use to evaluate $g_\eff$ in Figs.~\ref{fig:dt0}--\ref{fig:ar2test}.
The deviations from the measured pressure differences are comparable to their statistical uncertainty.

\refstepcounter{section}
\section*{Appendix \Alph{section}: Gravity-scan and averaging protocol}
\label{sec:protocol}
\setcounter{equation}{0}
\setcounter{figure}{0}

At fixed $\Delta T$, we change $g$ stepwise without reinitializing between neighboring values.
The principal scans use spacing $2\times10^{-5}$.
In Figs.~\ref{fig:dt0} and~\ref{fig:ar2test}(a), each point is $\langle X\rangle/L$, with the average taken over the latter half of the holding time at each $g$.

For the near-binodal system ($L/W=20$), the holding time is $2.5\times10^6$ Lennard--Jones time units for both temperature differences in Fig.~\ref{fig:dt0} and for the heat-conduction scans in Fig.~\ref{fig:intervals}; the $\Delta T=0$ reference in Fig.~\ref{fig:intervals} uses $5\times10^5$.
In Fig.~\ref{fig:dt0}, the equilibrium increasing and decreasing branches are prepared separately.
For $\Delta T=0.02$, the response in Fig.~\ref{fig:dt0} uses one run near the transition, $-4\times10^{-5}\leq g_\eff\leq6\times10^{-5}$, and another outside that range; it is not an average over runs.

For the transition points in Fig.~\ref{fig:intervals}, we classify the configuration at each sampled $g$ as liquid below (above) gas if at least $80\%$ of the late-time $X/L$ values lie below (above) $1/2$.
The sampled $g$ values across which this assignment changes bound the transition; an intermediate $g$ at which neither configuration meets this criterion enlarges that range.

For the hysteretic system ($L/W=2$), the holding time is $5\times10^5$ Lennard--Jones time units, and the repeated scans reverse direction at each endpoint.
Figure~\ref{fig:ar2test}(a) shows selected $g$ values from the first cycle of one run at each temperature difference.
Points for which the configuration changes during the holding time are omitted from Fig.~\ref{fig:ar2test}(a).

For the switching-frequency distributions in Fig.~\ref{fig:ar2test}(b), we separate the repeated scans into increasing and decreasing sweeps at the endpoint reversals.
Successive sweeps in the same direction are separated by a complete traversal of the sampled $g$ range.
For each temperature difference, 24 runs with two cycles each provide 48 switching events in each direction.
A switch is detected when five consecutive values of $X/L$ on the original side of $1/2$ are followed by five on the opposite side, with sampling interval $4\times10^3$ Lennard--Jones time units.
The switching value plotted is the $g_\eff$ at which this crossing occurs.
The frequencies count the first such switch in each directional sweep.

\refstepcounter{section}
\section*{Appendix \Alph{section}: Effective gravity in the free-energy landscape}
\label{sec:landscape}
\setcounter{equation}{0}
\setcounter{figure}{0}

In global thermodynamics, the liquid-below and liquid-above configurations are minima of a variational free energy \cite{NSgeff26,NSfull26}.
Let $F_{\rm LG}$ and $F_{\rm GL}$ denote their values.
To first order in weak gravity and heat conduction, the theory gives
\begin{align}
 F_{\rm GL}-F_{\rm LG}&\simeq2Nmg_\eff L\psi,
 \label{eq:landscape-minima}
\end{align}
where $\psi>0$ is the center-of-mass displacement $1/2-X/L$ of the liquid-below reference configuration at $g=\Delta T=0$.
The two minima therefore exchange their ordering at $g_\eff=0$.

Inside the spinodal region, the unstable single-phase state forms a ridge between the two minima.
Let $F_{\rm homo}$ denote its free energy, and define $\Delta F_0\equiv F_{\rm homo}-F_{\rm GL}=F_{\rm homo}-F_{\rm LG}$ at $g=\Delta T=0$.
The heights from the two minima are
\begin{align}
 F_{\rm homo}-F_{\rm LG}&\simeq\Delta F_0+Nmg_\eff L\psi, \nm
 F_{\rm homo}-F_{\rm GL}&\simeq\Delta F_0-Nmg_\eff L\psi.
 \label{eq:ridge-heights}
\end{align}
Both heights thus take their equilibrium form with $g$ replaced by $g_\eff$.
This gives a thermodynamic basis for comparing the hysteretic responses on the $g_\eff$ axis.

Between the binodal and spinodal, the single-phase state is instead metastable and may provide an intermediate route between the two separated configurations.
Nonequilibrium contributions not captured by $g_\eff$ enter these stationary free energies only at second order, but can change the surrounding landscape and transition paths \cite{NSfull26}.

\clearpage
\onecolumngrid
\section*{Supplemental Material: Movie descriptions}

Movies S1 and S2 show the reversal of the liquid--gas arrangement after heat conduction is introduced at fixed downward gravity.
Initially, both baths are at $\Tm=0.43$ and liquid lies below gas.
The lower bath is then heated to $T_1=0.44$ and the upper bath cooled to $T_2=0.42$, giving $\Delta T=0.02$ and $\ep_p=[\Ps(T_1)-\Ps(T_2)]/\Ps(\Tm)\simeq0.36$.
Liquid diminishes at the hot wall and grows at the cold wall without convective flow, producing liquid above gas.
Both movies show $2.5\times10^4$ Lennard--Jones time units before the temperature difference is imposed and $1.0\times10^5$ afterward.
All parameters are expressed in the Lennard--Jones units defined in the main text.

\paragraph*{Movie S1: $L/W=2$.}
The cell contains $N=3200$ particles at mean number density $\bar\rho=0.4$, with $g=1.32\times10^{-5}$.
Gravity points downward in the movie.
This is the same cell geometry and gravity as in Fig.~1(b),(c), but the movie uses the temperature difference $\Delta T=0.02$ specified above.
The supplied file is \texttt{MovieS1.mp4}.

\paragraph*{Movie S2: $L/W=20$.}
The elongated cell contains $N=3380$ particles at mean number density $\bar\rho=0.05$, with $g=2.0\times10^{-5}$.
The cell is displayed horizontally: gravity points right, the heated lower wall is on the right, and the cooled upper wall is on the left.
The thin liquid layer forms at the left wall against gravity, illustrating the configuration shown in Fig.~1(e).
The supplied file is \texttt{MovieS2.mp4}.

\end{document}